\documentclass[runningheads]{svmult}

\usepackage{makeidx}   
\usepackage[]{graphicx}  
\usepackage{subeqnar}  
\usepackage{multicol}  
\usepackage{physprbb}  
\makeindex             

\newcommand{\HI}   {H\,{\sc i}}
\newcommand{\HeII} {He\,{\sc ii}}
\newcommand{\HeI} {He\,{\sc i}}
\newcommand{\CIII} {C\,{\sc iii}}
\newcommand{\NIII} {N\,{\sc iii}}
\newcommand{\NV} {N\,{\sc v}}
\newcommand{\Halpha}{H$\alpha$}
\newcommand{\sun}{_{\odot}}
\begin{document}
\title*{Doppler Tomography of XTE\,J2123--058 and Other Neutron Star LMXBs}
\toctitle{Doppler Tomography of XTE\,J2123--058 and Other Neutron Star LMXBs}
%
%
\titlerunning{Doppler Tomography of Neutron Star LMXBs}
%
\author{Robert I. Hynes\inst{1}
\and Philip A. Charles\inst{1}
\and Carole A. Haswell\inst{2}
\and Jorge Casares\inst{3}
\and Cristina Zurita\inst{3}}
\authorrunning{Robert I. Hynes et al.}
%
%
\institute{
Department of Physics and Astronomy, University of Southampton, 
Southampton, SO17 1BJ, UK
\and 
Department of Physics and Astronomy, The Open University, Walton
Hall, Milton Keynes, MK7 6AA, UK
\and 
Instituto de Astrof\'\i{}sica de Canarias, 38200 La Laguna,
Tenerife, Spain}
\maketitle              
\begin{abstract}
We describe Doppler tomography obtained in the 1998 outburst of the
neutron star low mass X-ray binary (LMXB) XTE\,J2123--058.  This
analysis, and other aspects of phase-resolved spectroscopy, indicate
similarities to SW Sex systems, except that 
anomalous emission kinematics are seen in \HeII, whilst phase 0.5
absorption is confined to \Halpha.  This separation of these effects
may provide tighter constraints on models in the LMXB case than is
possible for SW Sex systems.  We will compare results for other LMXBs
which appear to show similar kinematics and discuss how models for the
SW Sex phenomenon can be adapted to these systems.  Finally we will
summarise the limited Doppler tomography performed on the class of
neutron star LMXBs as a whole, and discuss whether any common patterns
can yet be identified.
\end{abstract}
\section{Introduction}
Low mass X-ray binaries (LMXBs) contain a late-type `normal' star of
$\leq 1$\,M$_{\odot}$ accreting via
Roche lobe overflow onto a black hole or neutron star.  Black hole
systems are mostly only active during transient outbursts, making them
difficult targets for techniques requiring phase-resolved spectroscopy
such as Doppler tomography.  Many neutron star systems are
persistently active, but with a few exceptions are sufficiently faint
that they are still difficult targets.  Amongst the bright exceptions,
only Her X-1 has seen extensive application of Doppler tomography
\cite{Still:1997a,Quaintrell:2000a}. This is a very unusual system
with a companion star more massive than the neutron star, and not
strictly an LMXB at all.  We will focus here on the more limited data
available on `typical' LMXBs, i.e.\ those in which the companion star
is a late-type dwarf.  We begin by summarising our own work on a
transient neutron star system, XTE\,J2123--058 and then compare this
with other LMXBs.
\section{XTE\,J2123--058}
The transient LMXB XTE\,J2123--058 was discovered by {\it RXTE} on
1998 June 27 \cite{Levine:1998a} and promptly identified with a 17th
magnitude blue star with an optical spectrum typical of transients in
outburst \cite{Tomsick:1998a}.  Type-I X-ray bursts
\cite{Takeshima:1998a} indicated the compact object to be a neutron
star.  A pronounced photometric modulation \cite{Casares:1998a} on the
orbital period of 6.0\,hr
\cite{Tomsick:1998b,Ilovaisky:1998a,Hynes:1998a} is due to the
changing aspect of the X-ray heated companion indicating a high
inclination system.  During the outburst we obtained extensive
photometry \cite{Zurita:2000a} and phase-resolved spectroscopy
\cite{Hynes:2000a}; this section will focus on the latter.
\subsection{The Available Data}
We observed XTE\,J2123--058 using the ISIS dual-beam spectrograph on
the 4.2-m William Herschel Telescope (WHT) on the nights of 1998 July
18--20.  Full details and discussion of the data analysis are given in
\cite{Hynes:2000a}.  Phase resolved data suitable for Doppler
tomography were obtained on July 19--20 (approximately one binary
orbit each night) with an unvignetted wavelength range of
$\sim$4000--6500\,\AA\ and spectral resolution 2.9--4.1\,\AA.
\subsection{Doppler Tomography}
\begin{figure}[t]
\begin{center}
\includegraphics[angle=90,width=1.0\textwidth]{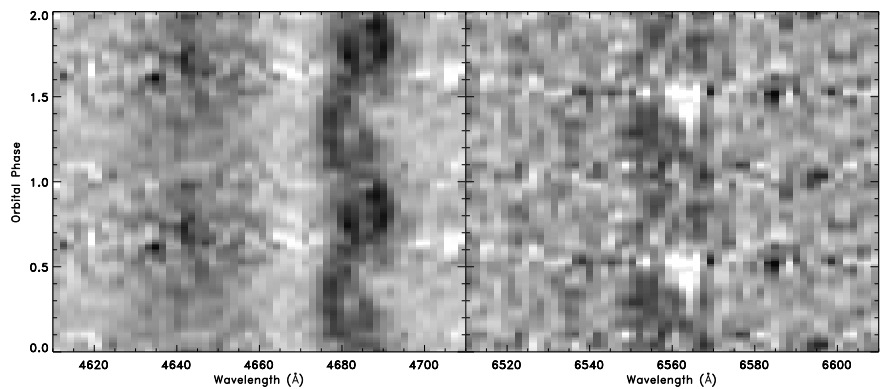}
\end{center}
\caption[]{Trailed spectrograms for XTE\,J2123--058.  The left hand
panel shows \CIII/\NIII\ and \HeII; the right \Halpha.  In the latter,
the transient absorption can be seen as the white region.}
\label{AbsFig}
\end{figure}

The emission lines show significant changes over an orbital cycle
(Fig.\ \ref{AbsFig}).  \HeII\ 4686\,\AA\ shows complex changes in line
position and structure, with two S-waves apparently interweaving.  The
light curve reveals a strong peak near phase 0.75 and a weaker one
near 0.25.

\begin{figure}[t]
\begin{center}
{\bf a)} \includegraphics[angle=270,width=.337\textwidth]{fig2a.ps}
\begin{minipage}[t]{.5\textwidth}
\vspace{-9mm}
\hspace*{0.5mm}
\begin{minipage}[b]{12pt}{\bf b)}\\\rule[0mm]{0mm}{57mm}\end{minipage}
\includegraphics[width=1.1\textwidth]{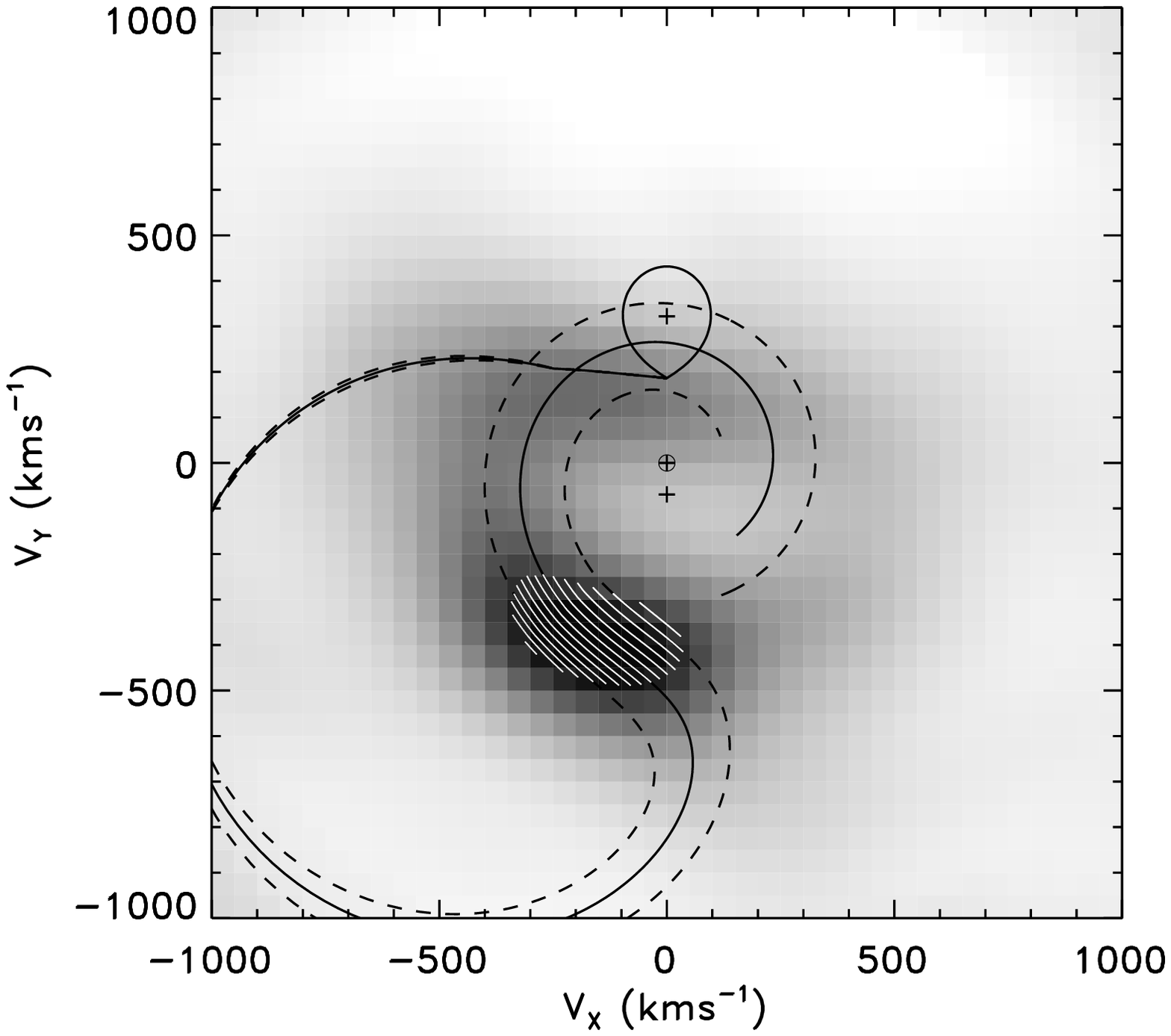}\\
\begin{minipage}[b]{12pt}{\bf c)}\\\rule[0mm]{0mm}{52mm}\end{minipage}
\includegraphics[width=\textwidth]{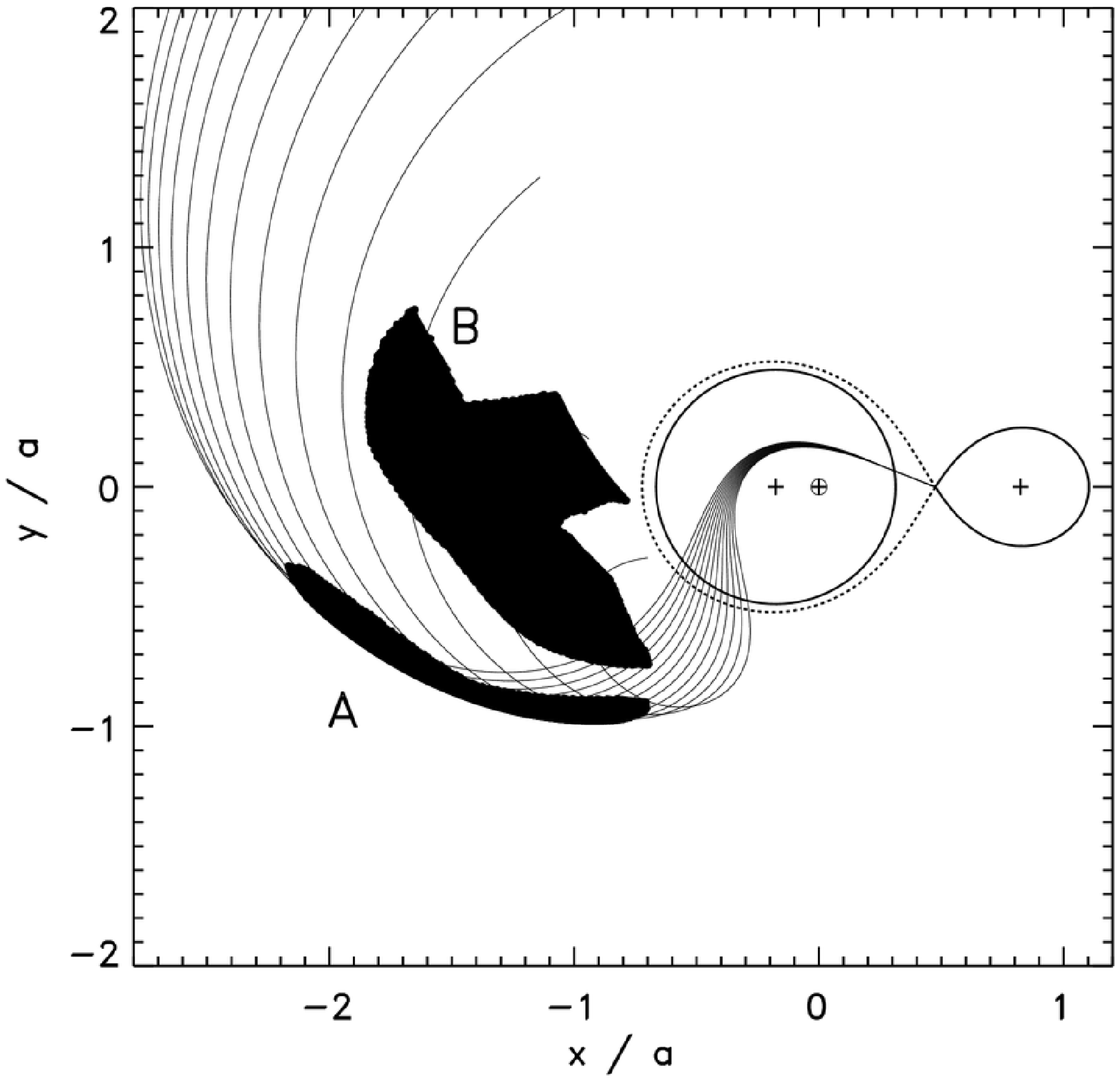}
\end{minipage}
\end{center}
\caption[]{Doppler tomography of XTE\,J2123--058.  a) From top to
bottom, the actual data, the tomogram and the reconstructed data.  In
the centre panel, the solid line is the ballistic stream trajectory,
the dashed line the Keplerian velocity at the stream position and the
large circle is the Keplerian velocity at the disc edge.  b)
Comparison between magnetic propeller trajectories and the
tomogram. c) Spatial plot corresponding to (b).  Points in region A
produce emission with kinematics corresponding to the white hatched
region in the tomogram.  Points in region B will produce absorption
with the phasing and velocities observed in \Halpha.}
\label{TomographyFig}
\end{figure}

We have used Doppler tomography \cite{Marsh:2000a} to identify \HeII\
4686\,\AA\ emission sites in velocity space (Fig.\
\ref{TomographyFig}a).  One of the fundamental assumptions of Doppler
tomography, that we always see all of the line flux at {\it some}
velocity, is clearly violated, however, as the integrated line flux is
not constant.  We attempt to account for this by normalising the line
profiles.  Without normalisation the structure of the derived tomogram
is very similar, so our results do not appear to be sensitive to this
difficulty.  A better solution would be to use the modulation mapping
method \cite{Steeghs:2000a}.

In order to interpret the tomogram quantitatively, we require
estimates of system parameters.  These were derived from results of
fits to outburst light curves \cite{Zurita:2000a}.  The parameters we
assume are $P_{\rm orb}=0.24821$\,day, $i=73^{\circ}$,
$M_1=1.4$\,M$_{\sun}$, $M_2=0.3$\,M$_{\sun}$, $R_{\rm
disc}=0.75\,R_{\rm L_1}$; see \cite{Hynes:2000a} for discussion.

The dominant emission site (corresponding to the main S-wave) appears
on the opposite side of the neutron star from the companion.  For any
system parameters, it is inconsistent with the heated face of the
companion and the stream/disc impact point.  There is a fainter spot
near the $\rm L_1$ point, corresponding to the fainter S-wave, which
is possibly associated with the accretion stream.  These spots in the
tomogram may be joined, although this may be a smoothing artifact of
the reconstruction.  It is notable that nearly all the emission has a
lower velocity than that of Keplerian material at the outer disc edge,
and so if associated with the disc requires sub-Keplerian velocities.
\subsection{A Neutron Star SW Sex System?}
The phenomenon of emission concentrated in the lower-left hand
quadrant of a Doppler tomogram at low velocities is not new; it is one
of the distinguishing characteristics of the SW~Sex class of
cataclysmic variables
\cite{Thorstensen:1991a,Horne:1999a,Hellier:2000a}.  These are all
novalike variables, typically seen at a relatively high inclination.

The tomograms are not the only similarity.  SW~Sex systems exhibit
transient absorption lines which are strongest near phase 0.5 or
slightly earlier and are moderately blue shifted.  Transient
absorption near phase 0.5 is seen in \Halpha\ in XTE\,J2123--058
(Fig.\ \ref{AbsFig}).  It begins slightly redshifted and moves to the
blue, and can be clearly seen in a trailed spectrogram.  Examination
of the average spectrum for phases 0.35--0.55 confirms that this is
real absorption below the continuum level.
\section{Other Candidate SW Sex-like LMXBs}
\begin{figure}[t]
\begin{center}
\includegraphics[width=.6\textwidth]{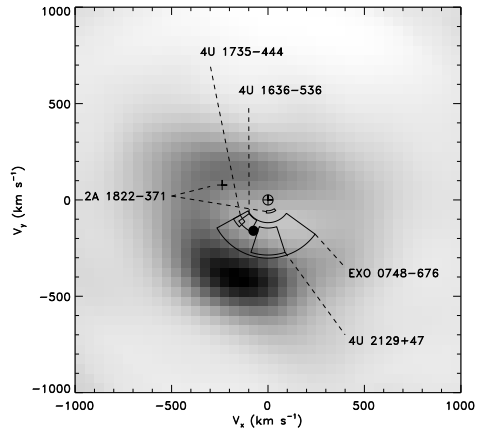}
\end{center}
\caption[]{Doppler tomography of XTE\,J2123--058 with radial velocity
information on other systems overlaid.  Boxes represent uncertainties
in velocity semi-amplitude and phasing.  The large point is the best
fit semi-amplitude and phase for the XTE\,J2123--058 data.}
\label{CompFig}
\end{figure}

The only other short-period neutron star LMXB with published Doppler
tomography is 2A\,1822--371 \cite{Harlaftis:1997a,Harlaftis:2000a}.
This was observed in \Halpha\ and exhibited disc emission enhanced
towards the stream impact point.  Similar behaviour is suggested by a
radial velocity analysis of \HeII\ \cite{Cowley:1982a}.  This is
different to what we see in XTE\,J2123--058.  Non-tomographic analysis
of other short-period neutron star LMXBs in an active state, however,
suggests a similar behaviour to that described for XTE\,J2123--058 as
shown in Fig.\ \ref{CompFig} where we overplot the \HeII\ radial
velocity information for other systems on our \HeII\ Doppler tomogram.
2A\,1822--371 is clearly the odd one out, and the other systems
(4U\,2129+47 \cite{Thorstensen:1982a}, EXO\,0748--676
\cite{Crampton:1986a}, 4U\,1636--536 \cite{Augusteijn:1998a} and
4U\,1735--444 \cite{Augusteijn:1998a}) all show emission centred in
the lower or lower-left part of the tomogram.


\section{SW Sex Models Applied to Neutron Star LMXBs}
Various models have been proposed for the SW~Sex phenomenon including
bipolar winds, magnetic accretion and variations on a stream overflow
theme.  Current opinion favours the latter interpretation, with the
stream either being accelerated out of the Roche lobe by a magnetic
propeller, or re-impacting the disc.  Both of these models are
discussed in more detail below.

\subsection{The magnetic propeller interpretation}
\label{PropellerSect}
It is possible to explain the behaviour of the \HeII\ 4686\,\AA\ line
in terms of the magnetic propeller model.  The essence of this model
is that in the presence of a rapidly rotating magnetic field,
accreting diamagnetic material may be accelerated tangentially,
leading to it being ejected from the system.  The field might be
anchored on the compact object, as in the prototype propeller AE~Aqr
\cite{Wynn:1997a,Eracleous:1996a} or in the disc
itself \cite{Horne:1999a}.

We adopt the parameterisation used by \cite{Wynn:1997a} and construct
a simple model of a propeller in XTE\,J2123--058.  Full details are
given in \cite{Hynes:2000a}.  We can readily find a trajectory which
passes through the central emission on the tomogram.  This is shown in
Fig.\ \ref{TomographyFig}b, together with two bracketing trajectories
corresponding to more and less acceleration.  For a plausible model,
we also require an explanation for why one particular place on the
trajectory is bright.  Such an explanation is offered by
\cite{Horne:1999a}: there is a point outside the binary at which
trajectories intersect.  At this point, faster moving blobs cross the
path of slower blobs and enhanced emission might be expected.  This
can be seen in Fig.\ \ref{TomographyFig}c.  To facilitate a
quantitative comparison of our data with the model, the region A in
Fig.\ \ref{TomographyFig}c encloses points with velocities consistent
with the bright emission spot.  These points are indeed located where
the trajectories cross.  Our data are therefore consistent with the
emission mechanism suggested by \cite{Horne:1999a}.  If the region in
which accelerated blobs collide is optically thick in the line then
emission is expected predominantly from the inner edge of this region
where fast moving blobs impact slow moving blobs.  For the geometry we
have considered, as can be seen in Fig.\ \ref{TomographyFig}c, this
will result in emission predominantly towards the top of the figure.
This corresponds to seeing maximum emission near phase 0.75, exactly
as is observed in our data.  Can we also explain the phase-dependent
Balmer absorption?  The region B in Fig.\ \ref{TomographyFig}c
encloses points which would absorb disc emission with the phasing and
velocities observed in \Halpha.  This region can be attributed to
lower-velocity trajectories which fall back towards the disc; thus the
transient absorption can be accommodated in this model.

In summary, a magnetic propeller model can account for both \HeII\
emission kinematics and lightcurve and \Halpha\ transient absorption.
This requires blobs to be ejected with a range of velocities;
absorption is caused by low velocity blobs, perhaps because these
account for most of the ejected mass.  Emission is only seen from
higher velocity blobs, which might be expected to exhibit more
energetic collisions.  The model has some difficulties, however, both for
a field anchored on the compact object and for a disc-anchored field.

If the field is anchored on the compact object then it must rotate
very fast, as spin periods of neutron stars are typically of order
milliseconds.  For such rapid rotation we must consider the
light-cylinder, defined by the radius at which the magnetic field must
rotate at the speed of light to remain synchronised with the compact
object spin \cite{Wynn:1997a}.  Outside of this radius,
$\sim6\times10^{-5}\,R_{\rm disc}$ for XTE\,J2123--058, the magnetic
field will be unable to keep up and hence becomes wound up.  This will
make the propeller less effective, although it may still produce some
acceleration (Wynn priv.\ comm., 2000).  The disc-anchored propeller
\cite{Horne:1999a} also has problems, in that it should very
efficiently remove angular momentum from the disc (Wynn \& King priv.\
comm., 1999), but this could perhaps be overcome if only a small
fraction of stream material passes through the propeller.  Finally,
when a propeller coexists with a disc, there is the added difficulty
in explaining how the accelerated material clears the disc rim.  In
the case of the disc-anchored propeller one can argue that field loops
emerging from the disc could accelerate material upwards as well as
out; for a propeller anchored on the neutron star some other
explanation would be needed.

\subsection{The stream overflow and disc re-impact interpretation}
The most popular alternative explanation for the SW Sex phenomenon is
the accretion stream overflow and re-impact model.  This was
originally suggested by \cite{Shafter:1988a}; for a recent exposition
see \cite{Hellier:2000a}.  This model has the advantage of being
physically very plausible.  Some stream overflow is implied by the
observations of X-ray dips in some LMXBs, and overflowing material
should re-impact in the inner disc if not supported in some way.  To
explain the observations, however, requires a number of additional
elements.

In this model the overflowing stream can be thought of in two regions.
The initial part of the stream produces the transient absorption when
seen against the brighter background of the disc.  The latter part,
where the overflow re-impacts the disc, is seen in emission giving
rise to the high velocity component.  In the simplest form of the
model, the overflow stream should produce absorption at all phases; it
always obscures some of the disc.  This problem is overcome by
invoking a strongly flared disc so that the overflow stream is only
visible near phase 0.5.  In XTE\,J2123--058, however, the neutron star
is directly visible, so the disc area that can be obscured by a flare
is limited and it is harder to reproduce the depth and transience of
the absorption.  A further difficulty is that to produce emission at
the low velocities observed requires disc emission which is
sub-Keplerian by a significant amount.  This problem is not peculiar
to XTE\,J2123--058 but common to the SW~Sex class in general.
\section{Conclusions}

\begin{table}[t]
\caption{Doppler tomography of neutron star LMXBs.  Bowen refers to
the \CIII/\NIII\ blend near 4640\,\AA.  Cen~X-4 was observed in
quiescence; the others in an X-ray active state.  Private
communications are (a) M. Torres, 2000, (b) Steeghs \& Casares, 2000}
\begin{center}
\setlength\tabcolsep{5pt}
\begin{tabular}{lllllll}
\hline\noalign{\smallskip}
& P$_{\rm orb}$ & Companion & Disc & Stream/ & Other & Source \\
& (hr)              &           &      & Hotspot &       &        \\
\noalign{\smallskip}
\hline
\noalign{\smallskip}
2A\,1822--371       &  5.6 &     & \HI & \HI    &        & 
\cite{Harlaftis:1997a,Harlaftis:2000a} \\
XTE\,J2123--058     &  6.0 &     &     & \HeII? & \HeII  & 
\cite{Hynes:2000a} \\
Cen X-4             & 15   & \HI & \HI & \HI?   &        & a \\
AC211               & 17   &     &     & \HeII? & \HeII? & 
\cite{Torres:2000a} \\
Sco X-1             & 19   & Bowen & \HeII, Bowen & &    & b \\
GX\,349+2           & 22   &     & \HI &        &        &
\cite{Wachter:1997a} \\
Her X-1             & 41   & \HI, \HeI, & & & \HeII  & 
\cite{Still:1997a,Quaintrell:2000a} \\
                    &          & Bowen, \NV & & &        & 
                                    \\
\noalign{\smallskip}
\hline
\noalign{\smallskip}
\end{tabular}
\end{center}
\label{CompTable}
\end{table}

We have demonstrated that XTE\,J2123--058 at least, and likely some
other short-period neutron star LMXBs, show observational similarities
to SW~Sex cataclysmic variables.  These include a Doppler tomogram
with emission concentrated at low velocities in the lower left
quadrant and transient absorption around phase 0.5.  We have discussed
the application of two current SW~Sex models to LMXBs.  The stream
overflow and re-impact model is theoretically plausible, but does not
readily account for observations.  The magnetic propeller model can
easily explain the main observations, at least in XTE\,J2123--058, but
has yet to overcome several theoretical objections.  Whatever the
correct interpretation, it is clear that the LMXB--SW~Sex connection
is an intriguing and fruitful one that may enhance our understanding
of both types of systems.

So far we have concentrated on short-period `normal' LMXBs; systems
similar to XTE\,J2123--058.  In Table \ref{CompTable} we summarise all
the Doppler tomography of neutron star LMXBs known to us.  The sample
is clearly small and we include this as much to highlight the lack of
data compared to cataclysmic variables as to draw strong conclusions.
One can perhaps suggest that the companion star mainly tends to be
prominent only in the longer period systems, and that `unusual'
tomograms (the other category) are most often seen in \HeII, but
both of these conclusions require a larger sample for confirmation.
It can be hoped that increasing availability of large telescope time
will make such observations possible.
\section*{Acknowledgements}

RIH, PAC and CAH would like to acknowledge support from grant
F/00-180/A from the Leverhulme Trust.  The William Herschel Telescope
is operated on the island of La Palma by the Isaac Newton Group in the
Spanish Observatorio del Roque de los Muchachos of the Instituto de
Astrof\'\i{}sica de Canarias.

%

\end{document}